# Effect of perceived preprint effectiveness and research intensity on posting behaviour

Pablo Dorta-González [1,*] and María Isabel Dorta-González [2]

[1] Universidad de Las Palmas de Gran Canaria, TiDES Research Institute, Campus de Tafira, 35017 Las Palmas de Gran Canaria, Spain. E-mail: pablo.dorta@ulpgc.es ORCID: http://orcid.org/0000-0003-0494-2903

[2] Universidad de La Laguna, Departamento de Ingeniería Informática y de Sistemas, Avenida Astrofísico Francisco Sánchez s/n, 38271 La Laguna, Spain. E-mail: isadorta@ull.es

* Corresponding author

**Abstract**

Open science is increasingly recognised worldwide, with preprint posting emerging as a key strategy. This study explores the factors influencing researchers' adoption of preprint publication, particularly the perceived effectiveness of this practice and research intensity indicators such as publication and review frequency. Using open data from a comprehensive survey with 5,873 valid responses, we conducted regression analyses to control for demographic variables. Researchers' productivity, particularly the number of journal articles and books published, greatly influences the frequency of preprint deposits. The perception of the effectiveness of preprints follows this. Preprints are viewed positively in terms of early access to new research, but negatively in terms of early feedback. Demographic variables, such as gender and the type of organisation conducting the research, do not have a significant impact on the production of preprints when other factors are controlled for. However, the researcher's discipline, years of experience and geographical region generally have a moderate effect on the production of preprints. These findings highlight the motivations and barriers associated with preprint publication and provide insights into how researchers perceive the benefits and challenges of this practice within the broader context of open science.

**Introduction**

In academic publishing, a 'preprint' refers to a version of a scholarly or scientific paper that precedes formal peer review and publication in a journal. As the University of Oxford articulates, a preprint is 'a full draft of a research paper that is shared publicly before it has been peer reviewed' (Oxford Open Access, n.d.). These pre-publication manuscripts allow researchers to disseminate their findings rapidly, solicit early feedback from the scientific community, and establish priority. While preprints offer several benefits, it is crucial to distinguish them from peer-reviewed journal articles, which have undergone a rigorous evaluation process to ensure their quality and validity.

Posting preprints is an important open science practice that accelerates scholarly publication and increases transparency (Ni & Waltman, 2024). This is because preprint servers allow researchers to make their work publicly available before it undergoes peer review (Hu et al., 2015). Over the past three decades, the prevalence of pre-printing has increased significantly (Xie et al., 2021). Today, publication on preprint servers is routine in several disciplines, including physics and mathematics (Brown, 2001; Larivière et al., 2014; Puebla et al., 2021). However, in many other fields, pre-printing remains less common. The large-scale survey by Ni and Waltman (2024) shows strong preprint adoption in physics, astronomy, mathematics and computer science, moderate uptake in the life and health sciences, and low use in social sciences and humanities.

The adoption of pre-printing varies considerably from region to region. A recent survey found that the United States and Europe are leading the way in preprint adoption, with respondents from these regions reporting greater familiarity with and commitment to preprints than their counterparts elsewhere (Ni & Waltman, 2024). In the life sciences, Abdill et al. (2020) showed that the United States and the United Kingdom contribute a disproportionately large number of preprints to bioRxiv compared to other countries. This disparity could be due to factors such as different levels of awareness of pre-printing or differences in the implementation of open science policies. In addition, the unique characteristics of the scholarly publishing system in countries such as China (Wang et al., 2021; Hyland, 2023) may also influence the adoption rate of preprints. However, there is growing interest in preprints globally, as evidenced by the rise of regional preprint servers such as AfricArXiv, ChinaXiv, Jxiv and SciELO Preprints (Chaleplioglou & Koulouris, 2023).

Pre-printing offers several potential benefits to authors, readers and other stakeholders, including reviewers and editors. It facilitates the immediate publication of research, helping to avoid duplication of effort and steering researchers away from unproductive research paths (Puebla et al., 2021). In addition, pre-printing allows authors to receive rapid feedback on their work (Malički et al., 2021; Rzayeva et al., 2023) and to claim priority for their findings (Pulverer, 2016; Vale & Hyman, 2016). As a permanent, citable record, preprints serve as evidence of productivity, which is particularly valuable for early-career researchers or those who do not plan to publish in traditional journals (Vale, 2015; Kim et al., 2020; Malički et al., 2021). Preprints also help to attract early attention from readers and editors (Barsh et al., 2016; Barrett, 2018), potentially increasing the visibility of the work and leading to more citations (Fu & Hughey, 2019; Fraser et al., 2020; Dorta-González & Dorta-González, 2023). Furthermore, open access research findings are more easily integrated into policy discussions and decision-making processes, underlining the value of open access practices in influencing real-world outcomes (Dorta-González et al., 2024). In the survey conducted by Ni and Waltman (2024), respondents highlighted the free accessibility of preprints and the acceleration of research communication as the most important benefits of pre-printing.

Despite the benefits, several challenges may hinder the widespread adoption of pre-printing. Common concerns include the risk of being scooped, questions about the reliability and credibility of unreviewed work, the potential for premature media coverage, geographical disparities in adoption rates, and potential conflict with certain journal policies (Chiarelli et al., 2019; Sever et al., 2019; Puebla et al., 2021; Fraser et al., 2022; Smart, 2022; Blatch-Jones et al., 2023; Ng et al., 2023). Journal policies vary considerably between disciplines. For example, Klebel et al. (2020) found that while 91% of journals in the life and earth sciences allow pre-printing, only 45% of journals in the humanities do so.

A survey conducted by ASAPbio (2020) found that concerns about pre-printing were more pronounced among respondents who had never published a preprint than among those with experience of pre-printing. Similarly, in a recent survey by Ni and Waltman (2024), the most important concerns about pre-printing were identified: low reliability and credibility, sharing results before peer review, and the risk of premature media coverage. These concerns were particularly highlighted by respondents in the life and health sciences. To encourage pre-printing, respondents highlighted the importance of integrating preprints into journal submission workflows and providing recognition for authors who publish preprints.

One aspect that has not been thoroughly explored in the literature is the isolation of the effect of researchers' perceptions of the effectiveness of preprint publication. To address this gap, our study examines the factors that influence researchers' adoption of preprint publication, with a particular focus on the perceived effectiveness of this practice. Using open data from a comprehensive survey with 5,873 valid responses, we conducted regression analyses to control for demographic variables such as research experience, gender, geographical region, type of organisation and academic discipline, as well as research intensity indicators such as publication and review frequency.

In this article, we address the following research questions: How do researchers' perceptions of the effectiveness of preprint publication influence their adoption and publication behaviour? What roles do demographic factors and research intensity play in shaping these attitudes?

**Data**

In this study, we used open data from a survey conducted by Research Consulting and the Centre for Science and Technology Studies (CWTS) at Leiden University, focusing on researchers' attitudes towards innovative research practices such as open access publishing, preprint publication and open peer review, and the incentives needed to encourage these behaviours (Chiarelli et al., 2024). The survey was part of a broader global, multi-stakeholder consultation commissioned by cOAlition S to assess the readiness of the research community for the transformative changes proposed in the Towards Responsible Publishing (TRP) initiative.

The consultation, carried out between November 2023 and May 2024, gathered input from over 11,600 participants, including 11,145 researchers who responded to a global survey, 440 respondents to an initial feedback survey, 72 participants in focus groups, and 10 organisational feedback letters from low- and middle-income countries. This extensive outreach was designed to capture a wide range of national, regional and disciplinary perspectives, ensuring that the findings are representative of the global research community.

Note that in standard linear regression, all observations must have complete information for every predictor and the outcome, as the least-squares estimators cannot be computed in the presence of missing values. Consequently, when a dataset contains any empty cells, the most common remedy is listwise deletion (also termed "complete-case analysis"), whereby any record with a missing datum on any variable of interest is excluded from the analysis. This procedure ensures that the mathematical requirements of the regression algorithm are satisfied by retaining only those cases with no missing entries.

Therefore, for our analysis, following a data-cleaning process, we identified 5,873 valid observations for regression analysis on the variables of interest. These observations were selected after excluding incomplete or invalid records to ensure the reliability and completeness of the data used in our study. This screening process was essential to conduct a robust examination of the factors influencing researchers' adoption of preprints and to ensure the integrity and accuracy of our findings.

**Method**

In this study, we used an Ordinary Least Squares (OLS) linear regression model to analyse the variable 'Production of preprints in the last three years', level 5 of Q6 in the survey, which measures recent preprint publication activity.

The independent variables in the model included 'Perceived effectiveness of preprints', derived from question Q13 of the survey. This question was rated on a Likert scale from 1 to 5, with 1 indicating 'not effective', 2 indicating 'slightly effective', 3 indicating 'moderately effective', 4 indicating 'very effective' and 5 indicating 'extremely effective'.

The specific formulation of Q13 was: 'Preprint publishing can be used to expedite the dissemination of scientific knowledge. How effective do you think preprints are in the following areas? (1) Providing early access to new research; (2) Receiving early feedback on new research; (3) Enhancing research accessibility and visibility; (4) Accelerating academic discourse; (5) Increasing research transparency'.

Other independent variables were related to the productivity of research outputs and peer reviews: 'production of research outputs in the last three years' (Q6), categorised into 4 levels excluding preprints, which is the dependent variable; and 'production of peer review outputs in the last three years' (Q22).

Control variables included the following demographic aspects: 'Primary research field' (Q1), categorised into 14 different fields; 'Organisation' (Q2), with 9 categories; 'Region' (Q3), with 14 geographical regions; 'Research experience' (Q4), categorised into 7 levels; and 'Gender' (Q5), with 4 categories.

Descriptive statistics for the dataset are presented in Tables 1 and 2. Table 1 offers a comprehensive overview of the demographics and professional background of the survey participants, highlighting the diversity across various dimensions. The largest proportion of respondents is from Life Sciences (14.4%) and Social Sciences (13.6%), followed by Medical and Health Sciences (13.4%) and Engineering (10.9%). The smallest proportions are in Law (1.5%) and Mathematics (2.5%).

Researchers are affiliated majority of Universities or Colleges (73.8%). Research Institutes (11.1%) and Hospitals or Medical Schools (4.9%) are also significant categories. Other types of organisations, such as Governmental Organizations (2.6%), Industry/commercial Organisations (2%) and Non-Governmental Organizations (1.0%), have smaller representations. The largest number of respondents is from Europe (50.8%), followed by Asia (21.8%) and the Americas (19.7%). Africa (5.2%) and Oceania (2.5%) have the fewest respondents. The distribution by geographical regions is shown in Table 1.

Most respondents are men (73.1%), with women constituting 24.3%. A small proportion identify as other genders (0.7%), and a few prefer not to disclose their gender (1.9%). Researchers have 6-9 years (19.8%) or 10-14 years (20.7%) of experience, with 15-24 years (20.4%) also representing a significant portion. Fewer respondents have less than 3 years (6.0%) or more than 24 years (16.3%) of experience.

Table 2 provides descriptive statistics for various quantitative variables related to research output and perceptions of preprint effectiveness, based on 5,873 observations. Respondents reported producing between 0 and 40 preprints, with a mean of 2.2 and a standard deviation of 4.9, revealing that preprint production is generally low. In contrast, the mean number of journal articles produced was 8.9, with an identical standard deviation, reflecting greater variability in the production of journal articles. Conference proceedings had a mean of 4.1 and a standard deviation of 6.3. Book chapters had a mean of 1.6 and a standard deviation of 3.6, pointing to a lower frequency compared to other types of research output. The production of books and monographs had a mean of 0.7 and a standard deviation of 2.5, suggesting relatively infrequent production. Finally, the number of peer reviews conducted ranged from 0 to 50, with a mean of 13.3 and a standard deviation of 13.9, showing considerable variability in peer review activity.

The effectiveness of preprints is evaluated using a Likert scale ranging from 1 to 5, where 1 represents "not effective" and 5 represents "extremely effective." Respondents generally perceive preprints as very effective for providing early access to research, with an average score of 3.8. They view preprints as somewhat effective for receiving early feedback, evidenced by a mean score of 3.4. Preprints are also considered effective in enhancing research accessibility and visibility, with a mean score of 3.7. Additionally, they are seen as moderately effective for accelerating academic discourse and increasing research transparency, both scoring a mean of 3.5.

**Results**

Table 3 presents the correlation matrix for various research output measures and perceptions of preprint effectiveness. The production of journal articles shows a moderate positive correlation with preprint production (0.35), revealing that researchers who publish more journal articles are likely to produce more preprints. In contrast, the production of conference proceedings has a weaker positive correlation with preprints (0.19), showing a minor association. Similarly, the production of book chapters, books and monographs all show weak positive correlations with preprints (0.19 and 0.23, respectively), revealing only slight associations. The number of peer reviews also exhibits a modest positive correlation with preprint production (0.20), indicating a minor link between peer review activities and preprints.

Regarding perceptions of preprints, providing early access to new research shows a weak positive correlation with preprint production (0.18), revealing some influence. Enhancing research accessibility and visibility, accelerating academic discourse, and increasing research transparency all display weak positive correlations with preprints (0.15, 0.13, and 0.10, respectively), showing limited relationships. However, receiving early feedback on new research has a very weak positive correlation with preprint production (0.03), suggesting minimal impact.

Interestingly, as can be seen in Table 3, the most active scientists - those who produce and review the most articles - tend to rate the effectiveness of preprints lower, particularly when it comes to early feedback on new research.

Overall, the correlations reveal that while there are some associations between research outputs and perceptions of preprints with the number of preprints posted, these relationships are generally modest. The regression analysis for the variable number of preprints is detailed in Tables 4 and 5. The adjusted R-squared value is 0.268, indicating that approximately 26.8% of the variability in the number of preprints can be explained by the regression model.

In Table 4, the significance of the model is underlined by a very low p-value (<0.0001), indicating that the predictors included in the model have a statistically significant impact on the number of preprints. This analysis shows that the regression model provides a significant improvement in predicting the number of preprints compared to a model that uses only the mean value of preprints. The high F-statistic, coupled with the extremely low p-value, confirms that the predictors of the model are collectively effective in explaining the variation in preprint numbers.

Table 5 shows the parameter estimates from the OLS regression model where the production of preprints in the last three years is the dependent variable. Among the research outputs, the number of journal articles has a positive and statistically significant relationship with preprint production (coef = 0.155, p < 0.0001). This means that for each added journal article produced, the likelihood of publishing a preprint increases by 0.15, holding all other variables constant (ceteris paribus). Similarly, the number of conference proceedings and book chapters also shows positive associations with preprint production (coef = 0.044, p < 0.0001; coef = 0.047, p = 0.017; respectively), although the effect is less pronounced, about one-third of that observed for journal articles. The production of books and monographs in the last three years has a particularly strong positive relationship with preprints (coef = 0.322, p < 0.0001), showing that each additional book or monograph increases the likelihood of depositing a preprint by 0.32, again holding all other variables constant. However, peer reviews show a modest positive association (coef = 0.018, p = 0.000), indicating a smaller but still significant impact on preprint production.

In terms of perceived effectiveness of preprints, providing early access to new research is positively associated with preprint production (coef = 0.465, p < 0.0001). This suggests that for every additional point (in the Likert 1-5 scale) of perceived effectiveness in providing early access to new research, the likelihood of depositing preprints increases by 0.46, holding all other variables constant. On the other hand, receiving early feedback hurts preprint production (coef = -0.371, p < 0.0001), meaning that for each additional point in the perceived effectiveness of receiving early feedback, the likelihood of depositing a new preprint decreases by 0.37, holding all other variables constant. This apparent contradiction is related to Table 3, which shows that most active scientists - those who produce and review the most articles - tend to rate the effectiveness of preprints lower, particularly when it comes to early feedback on new research.

In addition, increasing the accessibility and visibility of research, accelerating academic discourse, and increasing the transparency of research all show positive associations with preprint production, although the effects vary (coef = 0.305, p = 0.000; coef = 0.267, p = 0.001; coef = 0.185, p = 0.004, respectively). The interpretation stays consistent across cases. Specifically, each added point on the 1-5 Likert scale for perceived effectiveness in these three aspects increases the likelihood of depositing preprints by the amount indicated by the coefficient, all other variables held constant.

Disciplines such as Physics & Astronomy (coef = 3.643, p < 0.0001), Mathematics (coef = 2.577, p < 0.0001), and Computer Science (coef = 0.892, p = 0.002) produce significantly more preprints than Social Sciences. On the other hand, fields such as Medical and Health Sciences (coef = -0.463, p = 0.046), Engineering (coef = -0.442, p = 0.055) and Arts and Humanities (coef = -0.402, p = 0.099) produce significantly fewer preprints than Social Sciences.

Regional factors show significant differences when Western Europe is used as the reference category. Several geographical regions show a significant negative association with the number of preprints, revealing that their preprint production is significantly lower than that of Western Europe. Regions with lower preprint production over the last three years ($p < 0.01$) include South Asia (coef = -1.496), East and Central Asia (coef = -1.430), South America (coef = -1.390), Eastern Europe (coef = -1.351), Southeast Asia (coef = -1.300), North Africa (coef = -1.215) and Southern Europe (coef = -0.874).

Research experience also plays an important role. Using researchers with more than 24 years of experience as the reference category, younger researchers generally produce significantly more preprints. Specifically, researchers with 15-24 years, 3-5 years and 10-14 years of experience show positive associations with preprint production (coef = 0.533, p = 0.004; coef = 0.479, p = 0.021; coef = 0.393, p = 0.036, respectively), indicating significantly higher preprint production in the last three years compared to researchers with more than 24 years of experience.

Gender differences are minimal, with women and other gender identities showing little significant effect on preprint production compared to men. Furthermore, no significant association is seen between preprint production and the type of organisation, with the sole exception of independent researchers (coef = 1.475, p < 0.0001), who show a positive association.

Table 5 also includes the standardised coefficients in the last column to illustrate their relative effect sizes. Thus, the largest effect on preprint production is seen for the number of journal articles, with a standardised coefficient $\beta = 0.280$. This reveals that an increase in the production of journal articles is strongly associated with a significant increase in preprint production. This is closely followed by the number of books and monographs, which also has a strong positive effect with a coefficient $\beta = 0.163$, the third largest effect. Both results show that traditional forms of scholarly output are highly influential in predicting preprint adoption. Note that the highest field effect $\beta = 0.191$ for Physics and Astronomy falls between the effects of journal articles and books & monographs and constitutes the second largest effect.

There are positive effects for the number of peer reviews (β = 0.050) and conference proceedings (β = 0.057), but these are less pronounced than the effects for journal articles, books and monographs. In these cases, the effect size is about three times smaller than for books and monographs. The number of book chapters has the smallest positive coefficient in the productivity variables (β = 0.035), showing a relatively small impact on preprint production.

Among the perceived effectiveness of preprints, the highest effects are seen in providing early access to new research (β = 0.103) and receiving early feedback on new research (β = -0.087). These effects are among the largest, ranking in positions four and five by effect size, only slightly smaller than those for the productivity of journal articles and books. Improving the accessibility and visibility of research (β = 0.066) and accelerating academic discourse (β = 0.061) also have moderate positive effects, although they are somewhat smaller. Increasing research transparency (β = 0.045) has the smallest positive effect in this category.

**Discussion**

The disciplinary differences seen in receptivity towards preprints are consistent with findings from previous studies (Puebla et al., 2021; Ni & Waltman, 2024) and can be attributed to a combination of epistemological, cultural, and structural factors within academic fields. Disciplines such as physics, mathematics, and economics have long-established traditions of preprint sharing, where the early dissemination of findings is both normatively accepted and institutionally supported. These fields often prioritise rapid knowledge exchange and operate within relatively cohesive scholarly communities that facilitate informal peer feedback before formal publication. In contrast, disciplines in the humanities and some areas of the social sciences tend to place greater emphasis on monograph publication and narrative argumentation, where premature sharing of work may be viewed as compromising originality or scholarly ownership. Additionally, concerns about misinterpretation or reputational risk may be more pronounced in applied fields, where research often intersects with policy or public discourse. These varying disciplinary logics help explain why engagement with preprints stays uneven across the academic landscape.

The moderate effect of years of experience and geographical region on preprint production may be understood through the interplay of academic socialisation, access to infrastructure, and evolving norms in scholarly communication. Researchers with more years of experience may be more embedded in traditional publishing cultures and institutional reward systems that prioritise peer-reviewed journal articles, potentially leading to greater caution towards adopting novel practices such as pre-printing. Conversely, early-career researchers, while often more open to innovation, may face barriers related to perceived risks or a lack of mentorship in navigating open dissemination (Sarabipour et al., 2019). Geographical variation, meanwhile, reflects disparities in digital infrastructure, institutional support for open science, and the influence of regional policy frameworks (Tennant et al., 2019). For instance, researchers in Europe and parts of North America may benefit from stronger mandates and funding incentives for open access practices, including preprints, whereas those in under-resourced regions may encounter limited awareness, technical challenges, or concerns about visibility and recognition. These factors help explain why the observed effects, while statistically significant, are not uniformly strong across the sample.

The study finds that researchers' productivity in traditional communication channels, particularly the number of journal articles and books published, has the most significant impact on the frequency of preprint deposits. This finding underlines the centrality of traditional academic output metrics in shaping researchers' preprint behaviour. It reveals that researchers who are more prolific in established publication formats are also more likely to engage with preprints, possibly viewing them as an added means of disseminating their research quickly and widely.

The study also highlights the importance of researchers' perceptions of the effectiveness of preprints. While preprints are widely seen as a means of providing early access to new research, their role in easing early feedback is negatively perceived. This duality of perception reveals that while preprints are valued for accelerating the dissemination of research, there are reservations about their usefulness in promoting constructive peer engagement. These findings are novel and reveal a nuanced understanding of the perceived benefits and limitations of preprints and contribute to the ongoing discourse on their role in the scholarly communication ecosystem.

The unexpected negative association between valuing early feedback and preprint posting suggests that the benefits of feedback are not universally perceived as positive within the research community. While early visibility and access to research are often cited as clear advantages of pre-printing, feedback may be viewed as a more ambiguous or even risky element. One possible explanation is a misalignment between the type of feedback researchers hope for and what they actually receive through preprint platforms. If the feedback lacks depth or constructiveness, it may diminish the perceived value of sharing early work.

Furthermore, researchers might be concerned about reputational risks associated with receiving critique on work that is not yet fully polished. Those who place high importance on early feedback may, paradoxically, prefer to delay public dissemination until the manuscript has undergone more rigorous internal or peer review. This highlights the need for a more nuanced understanding of how feedback is experienced in the preprint context. To support wider adoption, preprint platforms and institutions may consider developing more structured, reliable, and supportive feedback mechanisms that better align with researchers' expectations and reduce perceived vulnerabilities.

Future research could delve deeper into why early feedback is perceived negatively and how it impacts preprint behaviour. Surveys or interviews could provide qualitative insights into researchers' concerns and preferences. Comparing the impact of feedback across different disciplines or research fields might reveal whether the negative effect is consistent or varies by context.

**Implications, limitations and recommendations**

The results of this study have several important implications for understanding the evolving landscape of scholarly publishing and the role of preprints within it. First, the strong correlation between traditional publication productivity and preprint deposit challenges the notion that preprints serve primarily as an alternative for researchers who find it difficult to get published in traditional venues. Instead, it appears that preprints complement existing publication strategies, particularly for those who are already successful in traditional formats. This finding is crucial for stakeholders looking to promote the use of preprints, as it reveals that efforts to increase preprint adoption may need to be tailored differently for researchers at different stages of their careers and with different publication profiles.

The study also shows that demographic variables, such as gender and type of organisation, do not have a significant impact on preprint production when controlling for other factors. This finding is significant because it reveals that the adoption of preprints is largely driven by factors related to researchers' work and perceptions, rather than inherent demographic characteristics. However, the moderate effects of discipline, experience and geographical region on preprint production reveal that disciplinary norms and regional academic cultures still play a role in shaping researchers' engagement with preprints. These findings suggest that efforts to promote preprints may need to be sensitive to disciplinary and regional differences, recognising that a 'one size fits all' approach may not be effective.

While the open dissemination of preprints offers clear benefits in terms of accelerating scientific communication and fostering collaboration, it also raises important ethical considerations in the context of generative artificial intelligence (GAI). Specifically, the public availability of preprints may inadvertently facilitate the use of early-stage research data by GAI tools to generate derivative content that may be misinterpreted or misused, particularly when the findings have not yet undergone peer review. This risk underscores the need for a balanced approach: promoting transparency and accessibility while also developing strategies to mitigate potential misuse. Measures such as watermarking preprints, including clear disclaimers about their preliminary nature, and encouraging responsible use of GAI technologies can help preserve the integrity of the scientific discourse.

Although the study provides valuable insights, it is not without limitations. One limitation is the potential for self-selection bias, as the researchers who participated in the study may have been more inclined to use preprints than the general population of researchers. In addition, the study's reliance on self-reported data introduces the possibility of response bias, where participants may have over- or under-reported their productivity or perceptions of preprints. Future research could address these limitations by using more objective measures of research productivity and preprint use.

The findings of this study have important recommendations for science policy, particularly about the promotion and support of preprint repositories. Given that traditional publication productivity is a key determinant of preprint use, policymakers should consider how preprints can be more fully integrated into existing academic reward structures. This could include recognising preprint repositories in research evaluations or incentivising researchers to share their work via preprints. In addition, the mixed perceptions of preprints suggest the need for policies that address researchers' concerns about the feedback process. This could include developing mechanisms to ensure more constructive and rigorous peer engagement with preprints.

For researchers, especially those early in their careers, the study highlights the importance of understanding how preprints can complement traditional publishing efforts. Institutions and funders should take note of the moderate effects of disciplinary and regional differences on preprint use and tailor their support and guidance accordingly. For publishers and platform providers, the findings suggest opportunities to improve the value proposition of preprints by addressing concerns about the feedback process and creating more integrated workflows that link preprints with traditional publishing processes.

**Conclusion**

This study provides a novel and comprehensive examination of the factors influencing researchers' adoption of preprint publication within the broader context of open science. Based on a robust analysis of survey data from 5,873 researchers across a range of disciplines and geographical regions, this study reveals important insights into how traditional scholarly productivity and perceptions of preprints drive preprint publication behaviour.

We found that researchers' productivity in traditional channels, such as journal articles and books, is the strongest predictor of their engagement with preprints. This finding challenges the prevailing assumption that preprints primarily serve as an alternative for those who are less successful in traditional publication venues.

We also found that while preprints are valued for providing early access to research, they are viewed less favourably for easing early feedback. This nuanced perception highlights a potential barrier to preprint adoption that has been underexplored in the existing literature.

Although demographic variables such as gender and type of organisation were found to have negligible effects, the researcher's discipline, experience and geographical region exert a moderate influence, suggesting that preprint adoption is also shaped by academic culture and norms.

To conclude, this study makes a significant contribution to the field of academic librarianship by providing novel insights into the factors driving preprint adoption and their implications for the scholarly communication ecosystem. These findings may stimulate further research and policy discussion on how to enhance the role of preprints in the global movement towards open science.

Table 1. Demographic characteristics of the sample (size = 5873)

| Variable | Category | Frequency | % |
|---|---|---:|---:|
| Field of research | Arts and Humanities | 501 | 8.5 |
| | Behavioural Sciences | 210 | 3.6 |
| | Chemistry | 291 | 5.0 |
| | Computer Science | 335 | 5.7 |
| | Earth Sciences | 277 | 4.7 |
| | Economics and Management | 336 | 5.7 |
| | Engineering | 641 | 10.9 |
| | Law | 91 | 1.5 |
| | Life Sciences | 847 | 14.4 |
| | Mathematics | 144 | 2.5 |
| | Medical and Health Sciences | 786 | 13.4 |
| | Physics and Astronomy | 427 | 7.3 |
| | Social Sciences | 799 | 13.6 |
| | Other | 187 | 3.2 |
| Type of organisation | Governmental organisation | 153 | 2.6 |
| | Hospital or medical school | 285 | 4.9 |
| | Independent researcher | 169 | 2.9 |
| | Industry/commercial organisation | 119 | 2.0 |
| | Non-governmental organisation (NGO) | 56 | 1.0 |
| | Research funding organisation | 24 | 0.4 |
| | Research institute (not in a university or medical school) | 652 | 11.1 |
| | University or college | 4336 | 73.8 |
| | Other | 78 | 1.3 |
| Continent | Africa | 307 | 5.2 |
| | Americas | 1156 | 19.7 |
| | Asia | 1283 | 21.8 |
| | Europe | 2981 | 50.8 |
| | Oceania | 145 | 2.5 |
| Region | Central America and the Caribbean | 203 | 3.5 |
| | Eastern Europe | 503 | 8.6 |
| | Eastern and Central Asia | 250 | 4.3 |
| | Northern Africa | 117 | 2.0 |
| | Northern America | 480 | 8.2 |
| | Northern Europe | 570 | 9.7 |
| | Oceania | 145 | 2.5 |
| | South America | 473 | 8.1 |
| | South-eastern Asia | 291 | 5.0 |
| | Southern Asia | 439 | 7.5 |
| | Southern Europe | 762 | 13.0 |
| | Sub-Saharan Africa | 190 | 3.2 |
| | Western Asia | 303 | 5.2 |
| | Western Europe | 1146 | 19.5 |
| Research experience | Less than 3 years | 350 | 6.0 |
|  | 3-5 years | 979 | 16.7 |
|  | 6-9 years | 1163 | 19.8 |
|  | 10-14 years | 1215 | 20.7 |
|  | 15-24 years | 1196 | 20.4 |
|  | More than 24 years | 955 | 16.3 |
|  | Not applicable | 14 | 0.2 |
| Gender | Woman | 1426 | 24.3 |
|  | Man | 4293 | 73.1 |
|  | Other | 42 | 0.7 |
|  | Prefer not to say | 111 | 1.9 |

Table 2. Descriptive statistics

| Variable | Obs. | Min. | Max. | Mean | SD |
|---|---|---|---|---|---|
| Production of research outputs in the last three years | | | | | |
| Num preprints | 5873 | 0 | 40 | 2.2 | 4.94 |
| Num journal articles | 5873 | 0 | 40 | 8.9 | 8.94 |
| Num conference proceedings | 5873 | 0 | 40 | 4.1 | 6.33 |
| Num book chapters | 5873 | 0 | 40 | 1.6 | 3.65 |
| Num books and monographs | 5873 | 0 | 40 | 0.7 | 2.50 |
| Num peer reviews | 5873 | 0 | 50 | 13.3 | 13.90 |
| How effective do you think preprints are in the following areas?[1] | | | | | |
| Providing early access to new research | 5873 | 1 | 5 | 3.8 | 1.09 |
| Receiving early feedback on new research | 5873 | 1 | 5 | 3.4 | 1.16 |
| Enhancing research accessibility and visibility | 5873 | 1 | 5 | 3.7 | 1.08 |
| Accelerating academic discourse | 5873 | 1 | 5 | 3.5 | 1.13 |
| Increasing research transparency | 5873 | 1 | 5 | 3.5 | 1.21 |

Note. [1] Likert scale 1-5: not effective (1), slightly effective (2), moderately effective (3), very effective (4), extremely effective (5)

Table 3. Linear correlations of Person

| | Num journal articles | Num conf proceedings | Num book chapters | Num books and monographs | Num peer reviews | Early access to new research | Early feedback on new research | Enhancing research accessibility and visibility | Accelerating academic discourse | Increasing research transparency | Num preprints |
|---|---|---|---|---|---|---|---|---|---|---|---|
| Num journal articles | **1** | 0.38 | 0.30 | 0.19 | 0.49 | -0.05 | -0.12 | -0.09 | -0.09 | -0.08 | 0.35 |
| Num conf proceedings | 0.38 | **1** | 0.34 | 0.28 | 0.26 | -0.05 | 0.01 | -0.02 | -0.01 | 0.00 | 0.19 |
| Num book chapters | 0.30 | 0.34 | **1** | 0.56 | 0.24 | -0.10 | -0.02 | -0.06 | -0.03 | -0.03 | 0.19 |
| Num books and monographs | 0.19 | 0.28 | 0.56 | **1** | 0.15 | -0.05 | 0.04 | -0.01 | 0.03 | 0.02 | 0.23 |
| Num peer reviews | 0.49 | 0.26 | 0.24 | 0.15 | **1** | -0.08 | -0.17 | -0.11 | -0.13 | -0.13 | 0.20 |
| Early access to new research | -0.05 | -0.05 | -0.10 | -0.05 | -0.08 | **1** | 0.48 | 0.62 | 0.56 | 0.45 | 0.18 |
| Early feedback on new research | -0.12 | 0.01 | -0.02 | 0.04 | -0.17 | 0.48 | **1** | 0.60 | 0.65 | 0.57 | 0.03 |
| Enhancing research accessibility and visibility | -0.09 | -0.02 | -0.06 | -0.01 | -0.11 | 0.62 | 0.60 | **1** | 0.72 | 0.61 | 0.15 |
| Accelerating academic discourse | -0.09 | -0.01 | -0.03 | 0.03 | -0.13 | 0.56 | 0.65 | 0.72 | **1** | 0.65 | 0.13 |
| Increasing research transparency | -0.08 | 0.00 | -0.03 | 0.02 | -0.13 | 0.45 | 0.57 | 0.61 | 0.65 | **1** | 0.10 |
| **Num preprints** | 0.35 | 0.19 | 0.19 | 0.23 | 0.20 | 0.18 | 0.03 | 0.15 | 0.13 | 0.10 | **1** |

Table 4. Analysis of variance (ANOVA) for the variable Y = number of preprints, calculated against the model Y=Mean(Y).

|  | DF | Sum of Squares | Mean Square | F | Pr > F |
|---|---|---|---|---|---|
| Model | 53 | 39,453 | 744.4 | 41.625 | **<0.0001** |
| Error | 5819 | 104,064 | 17.8 | | |
| Total Corrected | 5872 | 143,518 | | | |

Table 5. Effect of the explanatory variables on the production of preprints (OLS regression model)

| | Coef. | Std. err. | t | Pr > \|t\| | 95% CI Lower | 95% CI Upper | Sig. | Std. coef. β | Main effect ranking |
|---|---|---|---|---|---|---|---|---|---|
| Production of research outputs in the last three years | | | | | | | | | |
| Num journal articles | 0.155 | 0.008 | 19.404 | **<0.0001** | 0.139 | 0.170 | *** | 0.280 | #1 |
| Num conference proceedings | 0.044 | 0.010 | 4.330 | **<0.0001** | 0.024 | 0.064 | *** | 0.057 | |
| Num book chapters | 0.047 | 0.020 | 2.397 | **0.017** | 0.009 | 0.086 | * | 0.035 | |
| Num books and monographs | 0.322 | 0.027 | 11.873 | **<0.0001** | 0.269 | 0.375 | *** | 0.163 | #3 |
| Num peer reviews | 0.018 | 0.005 | 3.642 | **0.000** | 0.008 | 0.027 | *** | 0.050 | |
| How effective do you think preprints are in the following areas?[1] | | | | | | | | | |
| Providing early access to new research | 0.465 | 0.070 | 6.666 | **<0.0001** | 0.328 | 0.602 | *** | 0.103 | #4 |
| Receiving early feedback on new research | -0.371 | 0.068 | -5.471 | **<0.0001** | -0.504 | -0.238 | *** | -0.087 | #5 |
| Enhancing research accessibility and visibility | 0.305 | 0.084 | 3.626 | **0.000** | 0.140 | 0.469 | *** | 0.066 | |
| Accelerating academic discourse | 0.267 | 0.082 | 3.269 | **0.001** | 0.107 | 0.427 | ** | 0.061 | |
| Increasing research transparency | 0.185 | 0.065 | 2.861 | **0.004** | 0.058 | 0.311 | ** | 0.045 | |
| Primary field of research (reference category: Social Sciences) | | | | | | | | | |
| Arts and Humanities | -0.402 | 0.243 | -1.652 | 0.099 | -0.879 | 0.075 | . | -0.023 | |
| Behavioural Sciences | 0.338 | 0.330 | 1.024 | 0.306 | -0.309 | 0.985 | | 0.013 | |
| Chemistry | -0.250 | 0.297 | -0.841 | 0.400 | -0.831 | 0.332 | | -0.011 | |
| Computer Science | 0.892 | 0.284 | 3.138 | **0.002** | 0.335 | 1.449 | ** | 0.042 | |
| Earth Sciences | 0.250 | 0.301 | 0.830 | 0.406 | -0.340 | 0.839 | | 0.011 | |
| Economics and Management | -0.274 | 0.279 | -0.984 | 0.325 | -0.820 | 0.272 | | -0.013 | |
| Engineering | -0.442 | 0.230 | -1.921 | 0.055 | -0.894 | 0.009 | . | -0.028 | |
| Law | 0.251 | 0.472 | 0.532 | 0.594 | -0.674 | 1.177 | | 0.006 | |
| Life Sciences | -0.089 | 0.216 | -0.410 | 0.682 | -0.512 | 0.335 | | -0.006 | |
| Mathematics | 2.577 | 0.388 | 6.641 | **<0.0001** | 1.816 | 3.338 | *** | 0.081 | |
| Medical and Health Sciences | -0.463 | 0.232 | -1.995 | **0.046** | -0.917 | -0.008 | * | -0.032 | |
| Physics and Astronomy | 3.643 | 0.267 | 13.647 | **<0.0001** | 3.119 | 4.166 | *** | 0.191 | #2 |
| Other | -0.221 | 0.347 | -0.638 | 0.524 | -0.902 | 0.459 | | -0.008 | |
| Type of organisation (reference category: University or college) | | | | | | | | | |
| Governmental organisation | 0.056 | 0.352 | 0.160 | 0.873 | -0.634 | 0.746 | | 0.002 | |
| Hospital or medical school | 0.028 | 0.290 | 0.096 | 0.924 | -0.541 | 0.597 | | 0.001 | |
| Independent researcher | 1.475 | 0.334 | 4.411 | **<0.0001** | 0.819 | 2.130 | *** | 0.050 | |
| Industry/commercial organisation | 0.565 | 0.397 | 1.421 | 0.155 | -0.214 | 1.344 | | 0.016 | |
| Non-governmental organisation (NGO) | 0.883 | 0.574 | 1.539 | 0.124 | -0.242 | 2.008 | | 0.017 | |
| Research funding organisation | 0.481 | 0.868 | 0.554 | 0.580 | -1.221 | 2.182 | | 0.006 | |
| Research institute (not in a university or medical school) | 0.122 | 0.183 | 0.667 | 0.505 | -0.237 | 0.481 | | 0.008 | |
| Other | -0.075 | 0.487 | -0.155 | 0.877 | -1.031 | 0.880 | | -0.002 | |
| Region of the organisation (reference category: Western Europe) | | | | | | | | | |
| Central America and the Caribbean | -0.698 | 0.324 | -2.156 | **0.031** | -1.333 | -0.063 | * | -0.026 | |
| Eastern Europe | -1.351 | 0.227 | -5.941 | **<0.0001** | -1.797 | -0.905 | *** | -0.076 | |
| Eastern and Central Asia | -1.430 | 0.295 | -4.851 | **<0.0001** | -2.008 | -0.852 | *** | -0.058 | |
| Northern Africa | -1.215 | 0.413 | -2.942 | **0.003** | -2.025 | -0.406 | ** | -0.034 | |

| | | | | | | | | |
|---|---|---|---|---|---|---|---|---|
| Northern America | -0.375 | 0.229 | -1.638 | 0.101 | -0.825 | 0.074 | | -0.021 |
| Northern Europe | -0.082 | 0.214 | -0.383 | 0.701 | -0.502 | 0.338 | | -0.005 |
| Oceania | -0.376 | 0.373 | -1.008 | 0.313 | -1.108 | 0.355 | | -0.012 |
| South America | -1.390 | 0.233 | -5.976 | **<0.0001** | -1.846 | -0.934 | *** | -0.077 |
| South-eastern Asia | -1.300 | 0.282 | -4.617 | **<0.0001** | -1.852 | -0.748 | *** | -0.057 |
| Southern Asia | -1.496 | 0.244 | -6.128 | **<0.0001** | -1.975 | -1.018 | *** | -0.080 |
| Southern Europe | -0.874 | 0.195 | -4.474 | **<0.0001** | -1.257 | -0.491 | *** | -0.059 |
| Sub-Saharan Africa | 1.274 | 4.241 | 0.300 | 0.764 | -7.040 | 9.587 | | 0.003 |
| Western Asia | -0.656 | 0.276 | -2.377 | **0.017** | -1.197 | -0.115 | * | -0.029 |
| Research experience (reference category: More than 24 years) | | | | | | | | |
| Less than 3 years | 0.400 | 0.284 | 1.409 | 0.159 | -0.157 | 0.957 | | 0.019 |
| 3-5 years | 0.479 | 0.208 | 2.300 | **0.021** | 0.071 | 0.887 | * | 0.036 |
| 6-9 years | 0.188 | 0.195 | 0.964 | 0.335 | -0.194 | 0.569 | | 0.015 |
| 10-14 years | 0.393 | 0.188 | 2.093 | **0.036** | 0.025 | 0.762 | * | 0.032 |
| 15-24 years | 0.533 | 0.186 | 2.863 | **0.004** | 0.168 | 0.897 | ** | 0.043 |
| Not applicable | 0.501 | 1.150 | 0.436 | 0.663 | -1.753 | 2.755 | | 0.005 |
| Gender (reference category: Man) | | | | | | | | |
| Woman | -0.095 | 0.134 | -0.706 | 0.480 | -0.358 | 0.169 | | -0.008 |
| Other | 0.597 | 0.660 | 0.905 | 0.366 | -0.697 | 1.891 | | 0.010 |
| Prefer not to say | 0.139 | 0.410 | 0.338 | 0.735 | -0.665 | 0.942 | | 0.004 |
| _cons | -2.978 | 0.340 | -8.763 | **<0.0001** | -3.645 | -2.312 | *** | |
| Number of obs. | 5873 | | | | | | | |
| F | 41.625 | | | | | | | |
| Prob > F | <0.0001 | | | | | | | |
| Adjusted R-squared | 0.268 | | | | | | | |

*Note. Signification codes: 0 < \*\*\* < 0.001 < \*\* < 0.01 < \* < 0.05 < . < 0.1; [1] Likert scale 1-5: not effective (1), slightly effective (2), moderately effective (3), very effective (4), extremely effective (5)*